\documentclass[nofootinbib,preprintnumbers]{revtex4}

\usepackage[english]{babel}
\usepackage{graphicx}
\usepackage{psfrag}
\usepackage{amsmath}
\usepackage{amssymb}
\usepackage{verbatim}

\newcommand{\be}{\begin{equation}}
\newcommand{\ee}{\end{equation}}
\newcommand{\bea}{\begin{eqnarray}}
\newcommand{\eea}{\end{eqnarray}}
\newcommand{\bean}{\begin{eqnarray*}}
\newcommand{\eean}{\end{eqnarray*}}

\renewcommand{\b}{\langle}
\newcommand{\ket}{\rangle}

\newcommand{\e}{{\rm e}}

\renewcommand{\d}{{\rm d}}
\newcommand{\cl}[1]{{\mathcal #1}}

\newcommand{\pa}{\partial}

\newcommand{\ts}{\textstyle}
\newcommand{\sst}{\scriptstyle}

\newcommand{\bR}{\mathbb{R}}

\newcommand{\clS}{\cl{S}}

\newcommand{\clF}{\cl{F}}

\newcommand{\eq}[1]{(\ref{#1})}
\renewcommand{\sec}[1]{sec.\ \ref{#1}}

\newcommand{\fig}[1]{Fig.\ \ref{#1}}

\newcommand{\tr}{{\rm tr}}

\newcommand{\pic}[4]
{
 \begin{figure}
 \begin{center}
 \includegraphics[height=#3]{#4}
 \end{center}
 \caption{\label{#1} #2}
 \end{figure}
}

\newtheorem{theorem}{Theorem}[section]

\newtheorem{proposition}[theorem]{Proposition}

\newenvironment{proof}[1][Proof]{\begin{trivlist}
\item[\hskip \labelsep {\bfseries #1}]}{\end{trivlist}}
\newenvironment{definition}[1][Definition]{\begin{trivlist}
\item[\hskip \labelsep {\bfseries #1}]}{\end{trivlist}}

\newcommand{\qed}{\nobreak \ifvmode \relax \else
      \ifdim\lastskip<1.5em \hskip-\lastskip
      \hskip1.5em plus0em minus0.5em \fi \nobreak
      \vrule height0.75em width0.5em depth0.25em\fi}

\newcommand{\sixj}[6]{
\left\{
\begin{array}{ccc}
#1 & #2 & #3 \\ 
#4 & #5 & #6
\end{array}
\right\}
}

\newcommand{\Ct}{\tilde{C}}

\newcommand{\kp}{\kappa'}

\newcommand{\kt}{\tilde{\kappa}}

\begin{document}

\title{An exact string representation of 3d SU(2) lattice Yang--Mills theory}
\author{Florian Conrady}
\email{conrady@gravity.psu.edu}
\affiliation{Institute for Gravitational Physics and Geometry, Physics Department, Penn State University, University Park, Pennsylvania, U.S.A}
\author{Igor Khavkine}
\email{ikhavkin@uwo.ca}
\affiliation{Department of Applied Mathematics, University of Western Ontario, London, Ontario, Canada}
\preprint{IGPG-07/6-8}

\begin{abstract}
We show that 3d SU(2) lattice Yang--Mills theory can be cast in the form of an exact string representation.
The derivation starts from the exact dual (or spin foam) representation of the lattice gauge theory.
We prove that every dual configuration (or spin foam) can be equivalently described as a self--avoiding worldsheet of strings on a framing of the lattice.
Using this correspondence, we translate the partition function into a sum over closed worldsheets that are weighted with explicit amplitudes.
The expectation value of two Polaykov loops with spin $j$ becomes a sum over worldsheets that are bounded by $2j$ strings along a framing 
of the loops.
\end{abstract}

\maketitle

\section{Introduction}
\label{introduction}

It has been a long--standing conjecture that gauge theory has a dual or effective description in terms of string--like degrees of freedom (see e.g.\ \cite{Polyakovgaugefieldsstrings,Polyakovliberation,Polyakovconfiningstrings,Antonovstringnature} for a review). The idea made its first appearance in the 60's when dual resonance models of hadron scattering were interpreted in terms of strings \cite{Nambuselectedpapers,SusskinddualsymmetricI,FryeLeeSusskind,FairlieNielsen,Schwarzearlyyears}. It reappeared again, when Wilson introduced the strong--coupling expansion and argued that confinement is an effect of flux lines between quarks \cite{Wilsonconfinement}. This motivated attempts to find an exact or effective string representation of Yang--Mills theory, and led to the study of loop equations \cite{Polyakovloopequations,Migdal}, and to lattice and continuum models of the Nambu--Goto string (see e.g.\ \cite{LuscherWeiszSymanzik,Alvarezstaticpotential,Arvis} and refs.\ in \cite{Ambjornbookquantumgeometry}). Then, critical string theory was introduced \cite{BrinkDiVecchiaHowelocally,BrinkDiVecchiaHoweLagrangian,Polyakovbosonicstring}, and developed further into superstring theory, with the broader aim of unifying gauge theory, matter and gravity. Another aspect of the gauge--string duality was revealed when 't Hooft analyzed the large $N$ limit of perturbative Yang--Mills theory \cite{tHooftlargeNlimit}. In the context of string theory, the idea was revived more recently by the AdS--CFT correspondence: by the conjecture that a supersymmetric conformal Yang--Mills theory has an equivalent description in terms of superstrings in an AdS spacetime \cite{Maldacena,GubserKlebanov,Wittenholography,AharonyGubserMaldacenaOoguriOz}. 

Conceptually, the present paper is close to Wilson's original approach, where flux lines arise as diagrams of a strong--coupling expansion.
There are different versions of the strong--coupling expansion that have different convergence properties. 
Here, we are concerned with the ``resummed'' expansion that is convergent for any coupling \cite{Munsterhightemperature,DrouffeZuber}:
it results from an expansion of plaquette actions into a basis of characters, and from a subsequent integration over the connection. 
Thus, the sum over graphs is not an expansion in powers of $\beta$, but rather a \textit{dual representation} that is equivalent to the original lattice gauge theory \cite{Anishettyetal,HallidaySuranyi,DiakonovPetrov,OecklPfeifferdualofpurenonAbelian}. For this reason, we try to avoid the adjective ``strong--coupling'' and call the graphs instead \textit{spin foams} \cite{OecklPfeifferdualofpurenonAbelian}. Originally, this name was introduced for SU(2) \cite{Baezspinfoammodels}, but it is also used for general gauge groups. In the case of SU(2), one obtains a sum over spin assignments to the lattice that satisfy certain spin coupling conditions. Each admissible configuration is a spin foam.

To some extent, the concept of spin foams already embodies the idea of an exact gauge--string duality: spin foams can be considered as branched surfaces that are worldsheets of flux lines (see sec.\ 6.3 in \cite{ItzyksonDrouffestatistical} and \cite{Conradygeometricspinfoams}). Due to the branching and the labelling with representations, these surfaces are not worldsheets as in string theory, however. 

The new element of this paper is the following: we show that in 3 dimensions spin foams of SU(2) can be decomposed into worldsheets that do not branch and carry no representation label. They can be regarded as worldsheets of strings in the fundamental representation. To carry out this decomposition, we have to apply two modifications to the lattice: the cubic lattice is replaced by a tesselation by cubes and truncated rhombic dodecahedra. This ensures that at every edge exactly three faces intersect. In the second step, the 2--skeleton of this lattice is framed (or thickened). The thickening allows us to replace each spin assignment $j_f$ to a face by $2j_f$ sheets of a surface. We show that these sheets can be connected to form a worldsheet in the thickened complex. Moreover, by imposing suitable restrictions on the worldsheets, we can establish a bijection between spin foams and worldsheets.
 
Once this bijection is given, it is simple to rewrite exact sums over spin foams as exact sums over worldsheets. The boundary conditions depend on the observable that is computed by the spin foam sum. In the case of a Wilson loop in the representation $j$, the sum extends over worldsheets that are bounded by $2j$ closed strings. In this paper, we derive the sum over worldsheets explicitly for two Polyakov loops of spin $j$ that run parallel through the lattice.
 
The paper is organized as follows: in section \ref{spinfoamsandspinnetworks} we set our conventions for spin foams and their boundaries (so--called spin networks).
Then, we specify 3d SU(2) lattice Yang--Mills theory with the heat kernel action (\sec{SU2latticegaugetheory}). In section \ref{spinfoamrepresentation}, we describe the dual transform of the partition function and of the expectation value of two Polyakov loops. The central part of the paper is section \ref{worldsheetinterpretationofspinfoams}, where we introduce worldsheets on the framed lattice, and prove the bijection between worldsheets and spin foams.
In the final section, we formulate  both the partition function and the expectation value of the Polyakov loops as exact sums over worldsheets with explicit amplitude factors.

\section{Spin foams and spin networks}
\label{spinfoamsandspinnetworks}

In this section, we set our conventions for spin foams and spin networks of SU(2). Spin networks formalize the concept of flux line, and spin foams can be regarded as worldsheets of these flux lines. 

In this paper, spin foams will live on 3--complexes where at each interior edge exactly three faces meet. Spin networks will only lie on the boundary of this complex. For this reason, we do not need to consider the most general concept of spin foam and spin network that could occur and restrict ourselves to the following definition.

Let $\Lambda$ be a complex where at each interior edge exactly three faces meet. A spin foam $F$ on $\Lambda$ is given by an assignment of a spin $j_f$ to every face $f$ of $\Lambda$ such that at every interior edge $e$ of $\Lambda$ the triangle inequality is satisfied by the three adjacent spins.
Dually, the spin foam can be described as a configuration on the dual complex $\Lambda^*$: then, the spin foam $F$ is specified by spins $j_e$ on edges of $\Lambda^*$, where for every triangle of $\Lambda^*$, the spins on the edges of the triangle satisfy the triangle inequality.

We define a spin network $S$ on the boundary $\pa\Lambda$ as an assignment of spins $j_e$ to edges in the boundary $\pa\Lambda$ such that for every vertex in the boundary the adjacent spins satisfy the triangle inequality. A particularly simple example of a spin network is a non--selfintersecting loop $C$ that carries a spin label $j$. We denote such a spin network by $(C,j)$. Each spin foam on $\Lambda$ induces a spin network on the boundary $\pa\Lambda$, which we call the boundary $\pa F$ of $F$.

\section{SU(2) lattice Yang-Mills theory in 3 dimensions}
\label{SU2latticegaugetheory}

The partition function of 3--dimensional SU(2) lattice Yang--Mills theory is defined by a path integral over SU(2)-valued link (or edge) variables $U_e$ on a cubic  lattice $\kappa$:
\be
\label{partitionfunction}
Z = \int\left({\ts\prod\limits_{e\subset\kappa}}\d U_e\right) \exp\Big(-\sum_f \clS_f(U_f)\Big)
\ee
The face (or plaquette) action $\clS_f$ depends on the holonomy $U_f$ around the face. As in paper I, we choose $S_f$ to be the heat kernel action (for more details on the definition, see \cite{MenottiOnofri}). The heat kernel action has a particularly simple expansion in terms of characters, namely, 
\be
\exp\Big(- \clS_f(U_f)\Big) = \sum_j\;(2j+1)\,\e^{-\frac{2}{\beta}\,j(j + 1)}\,\chi_j(U_f)\,.
\ee
The coupling factor $\beta$ is related to the gauge coupling $g$ via
\be
\beta = \frac{4}{a g^2} + \frac{1}{3}\,.
\ee
The expectation value of a Wilson loop $C$ in the representation $j$ is
\be
\label{Wilsonloop}
\b \tr_j U_C\ket = \int\left({\ts\prod\limits_{e\subset\kappa}}\d U_e\right)\; \tr_j U_C\,\exp\Big(-\sum_f \clS_f(U_f)\Big)\,.
\ee
$U_C$ denotes the holonomy along the loop $C$.

\section{Spin foam representation}
\label{spinfoamrepresentation}

\subsection{Partition function}

\psfrag{j1}{$\sst j_1$}
\psfrag{j2}{$\sst j_2$}
\psfrag{j3}{$\sst j_3$}
\psfrag{j4}{$\sst j_4$}
\psfrag{j5}{$\sst j_5$}
\psfrag{j6}{$\sst j_6$}

\pic{TandTdual}{Tesselation of $\bR^3$ by cubes and truncated rhombic dodecahedra.}{7cm}{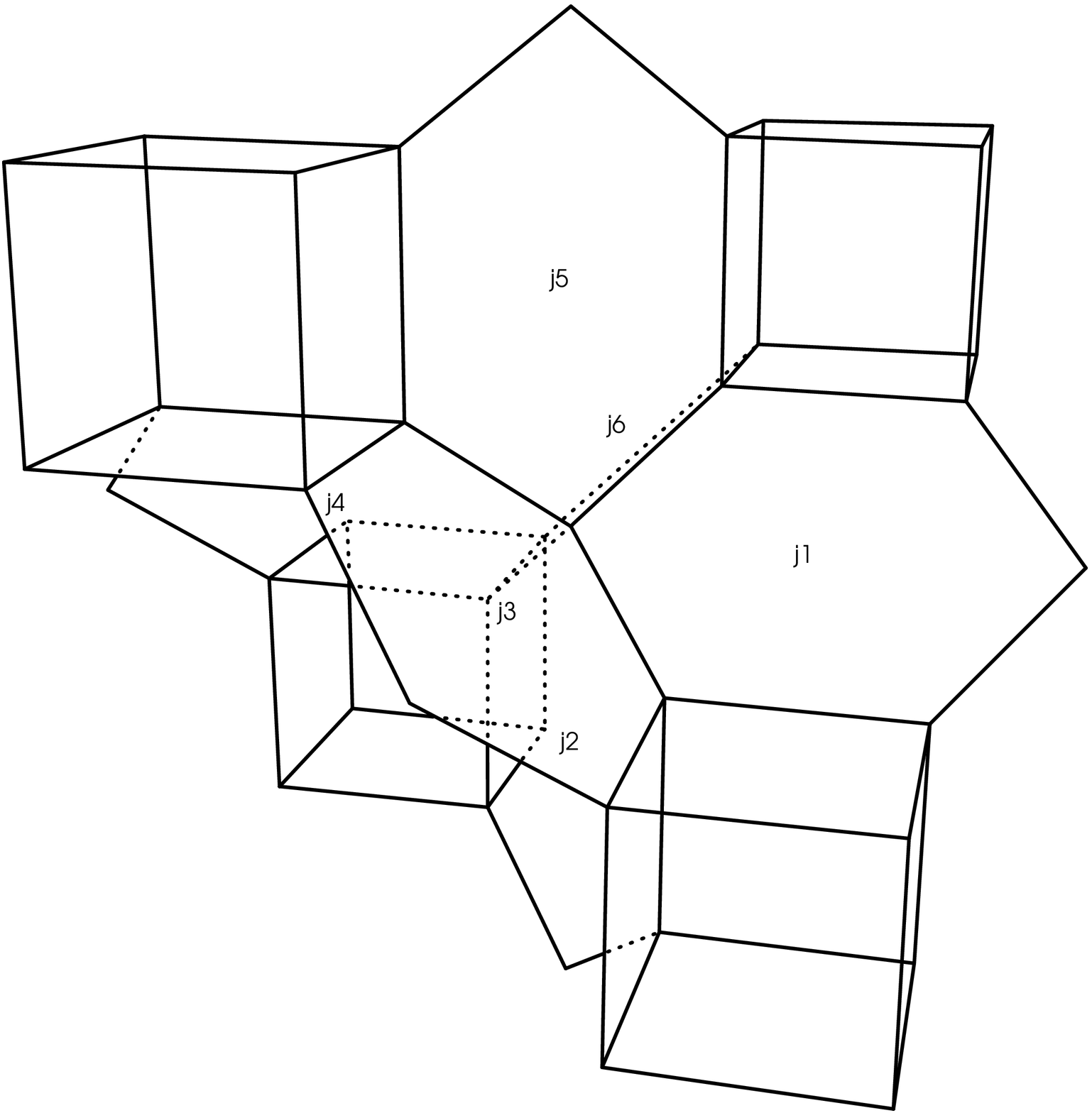}

In general, there are several, equivalent ways of writing down a sum over spin foams. Here, we will use a scheme by Anishetty, Cheluvaraja, Sharatchandra and Mathur \cite{Anishettyetal}, where the amplitude is expressed in terms of $6j$--symbols\footnote{Recently, the same result was obtained very efficiently by the use of Kauffman--Lins spin networks \cite{ChristensenCherringtonKhavkine}.}. In the paper by Anishetty et al., spin foams are described by spin assignments $j_e$ to \textit{edges} of a triangulation $T$. For the purpose of the present paper, it is convenient to go to the dual picture where spin foams are spin assignments $j_f$ to \textit{faces} of the dual $T^*$. Let us call this lattice $\kt$. It is given by a tesselation of the 3--dimensional space by cubes and truncated rhombic dodecahedra (see \fig{TandTdual}).

The complex $\kt$ contains two types of faces: square faces that correspond to faces of the original cubic lattice $\kappa$, and hexagonal faces that connect pairs of square faces. At each edge of $\kt$, exactly three faces meet, and at each vertex we have six intersecting faces. We will be slightly sloppy with our notation and write $f\subset\kappa$ to denote the square faces of $\kt$. 

After the dual transformation, the partition function \eq{partitionfunction} is expressed as a sum over closed spin foams $F$ on $\kt$, where each spin foam carries a certain weight:
\be
\label{spinfoamsumpartitionfunction}
Z = \sum_{F\;|\;\pa F = \emptyset}
\left(\prod_{f\subset\kt} (2j_f+1)\right)
\left(\prod_{v\subset\kt} A_v\right)
\left(\prod_{f\subset\kappa}\;(-1)^{2j_f}\,\e^{-\frac{2}{\beta}\,j_f(j_f + 1)}\right)\,.
\ee
In the amplitude, every face contributes with the dimension $2j_f + 1$ of the representation $j_f$. In addition, square faces give an exponential of the Casimir and a sign factor $(-1)^{2j_f}$. For each vertex of $\kt$, we get the value of a so-called tetrahedral spin network as a factor:
\be
A_v \quad=\quad \parbox{2.9cm}{\includegraphics[height=2.8cm]{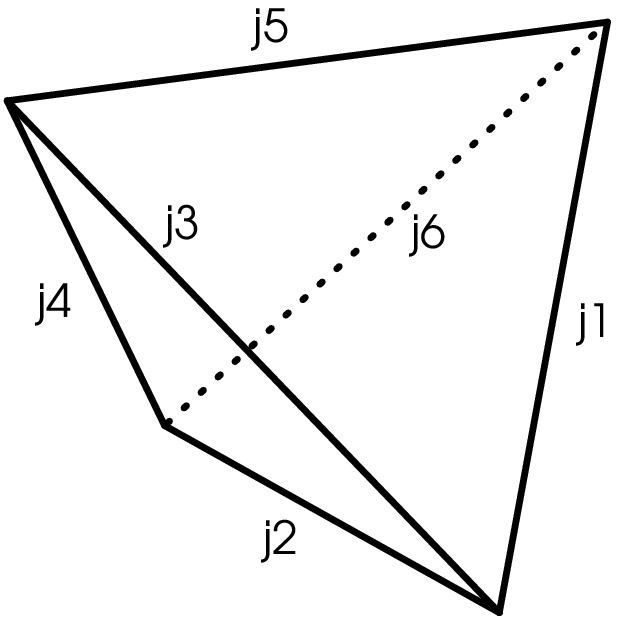}} \quad=\quad \sixj{j_1}{j_2}{j_3}{j_4}{j_5}{j_6} 
\ee
The edges of the tetrahedral spin network correspond to faces of the spin foam surrounding the vertex $v$, and the vertices of the spin network correspond to the edges where these faces meet (see \fig{TandTdual}). The value of the spin network is equal to a $6j$-symbol, where the spins $j_1$, $j_2$ and $j_3$ are read off from any vertex of the tetrahedron.

\subsection{Polaykov loops}
\label{subsectionPolyakovloops}

The dual transformation can be also applied to expectation values of observables such as Wilson loops or products of them.
When the dual transform of such loops is computed, the explicit form of amplitudes depends on the geometry of the loops.
For a rectangular Wilson loop, it was explicitly determined by Diakonov \& Petrov \cite{DiakonovPetrov}.
In ref. \cite{ConradydualPolyakovloop}, one of us derived the dual amplitude for Polyakov loops. 
In the following, we will consider the example of Polyakov loops, since everywhere along the loops the amplitude has the same structure. 
In the case of a rectangular Wilson loop, one has to distinguish between the straight part and the corners of the loop.

\psfrag{C1}{$C_1$}
\psfrag{C2}{$C_2$}
\pic{polyakovloop}{Zig--zag path of the Polyakov loops $C_1$ and $C_2$ in a 2d slice of the lattice $\kappa$. The arrows indicate how lattice points are identified.}{6cm}{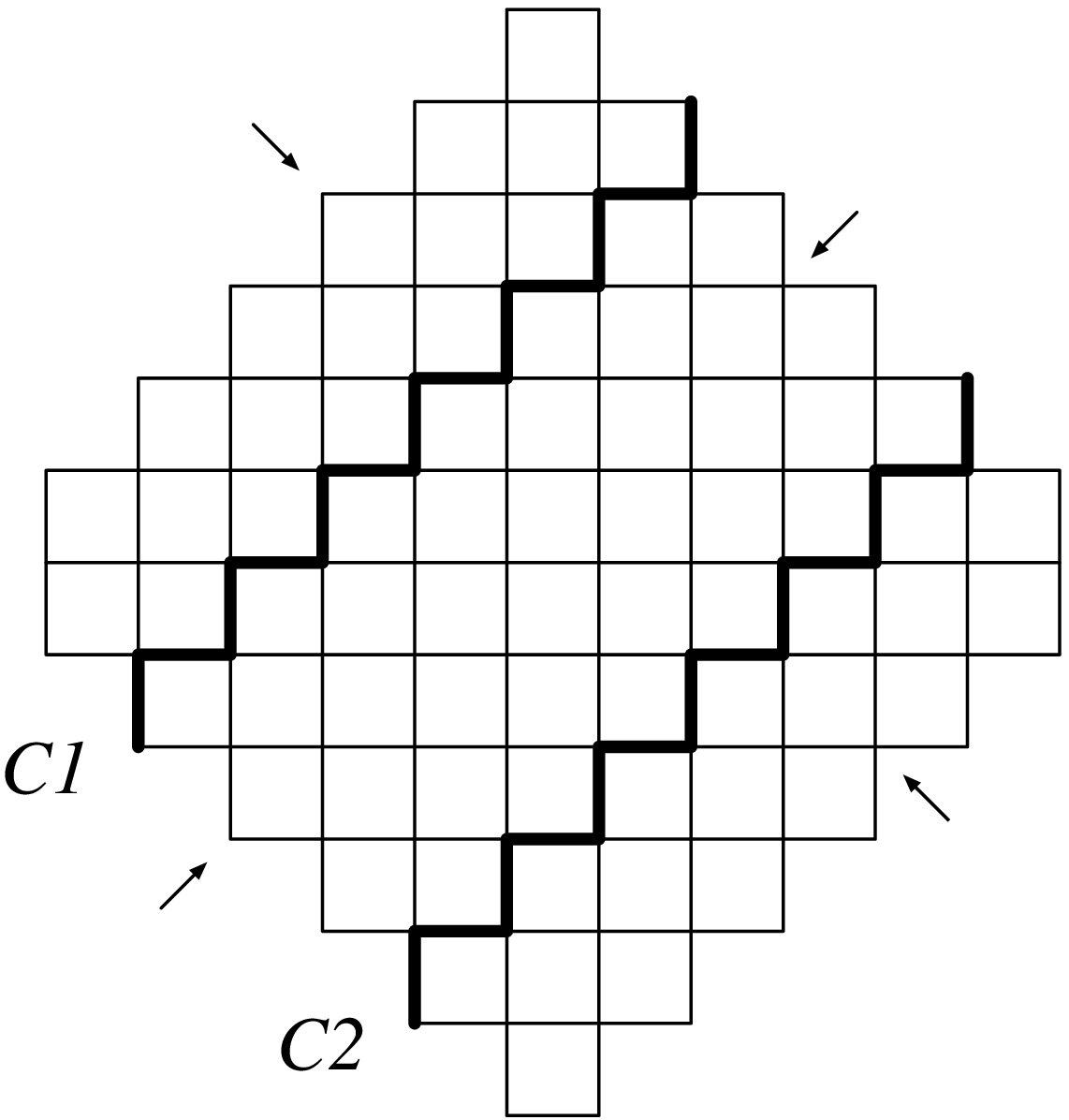}

We let the Polyakov loops $C_1$ and $C_2$ run along zig--zag paths through the lattice $\kappa$ and adopt boundary conditions that identify lattice points on opposing ends of diagonals\footnote{The use of zig--zag paths is not essential for the result of this paper. We choose these paths for convenience, since in this case the amplitudes are already known from ref.\ \cite{ConradydualPolyakovloop}.} (see \fig{polyakovloop}). As before, we introduce a tesselation $\kt$, where square faces correspond to faces of the original lattice, and hexagonal faces connect pairs of such faces. To describe the spin foam sum for the Polyakov loops, we need to modify this lattice. This happens in several steps: first we remove all 3--cells, so that we obtain the 2--skeleton of $\kt$. In $\kt$ the Polyakov loops $C_1$ and $C_2$ correspond to two closed sequences of hexagons. Imagine that we draw a closed loop within each sequence that connects the centers of neighbouring hexagons (see \fig{Tdualprime}). For each pair of neighbouring hexagons, we also add an edge that connects their centers directly, i.e.\ in a straight line outside the 2--complex. Each such edge forms a triangle with the edges inside the hexagons. We include these triangular faces in the complex, and call the resulting 2--complex again $\kt$. Its boundary consists of two loops which we denote by $\Ct_1$ and $\Ct_2$ respectively.

\psfrag{j1}{$\sst j_1$}
\psfrag{j2}{$\sst j_2$}
\psfrag{j3}{$\sst j_3$}
\psfrag{j4}{$\sst j_4$}
\psfrag{j5}{$\sst j_5$}
\psfrag{j6}{$\sst j_6$}
\psfrag{j1'}{$\sst j'_1$}
\psfrag{j3'}{$\sst j'_3$}
\psfrag{j}{$\sst j$}
\pic{Tdualprime}{Modification of the complex $\kt$: the effect of the Polyakov loops can be described by inserting additional faces.}{6cm}{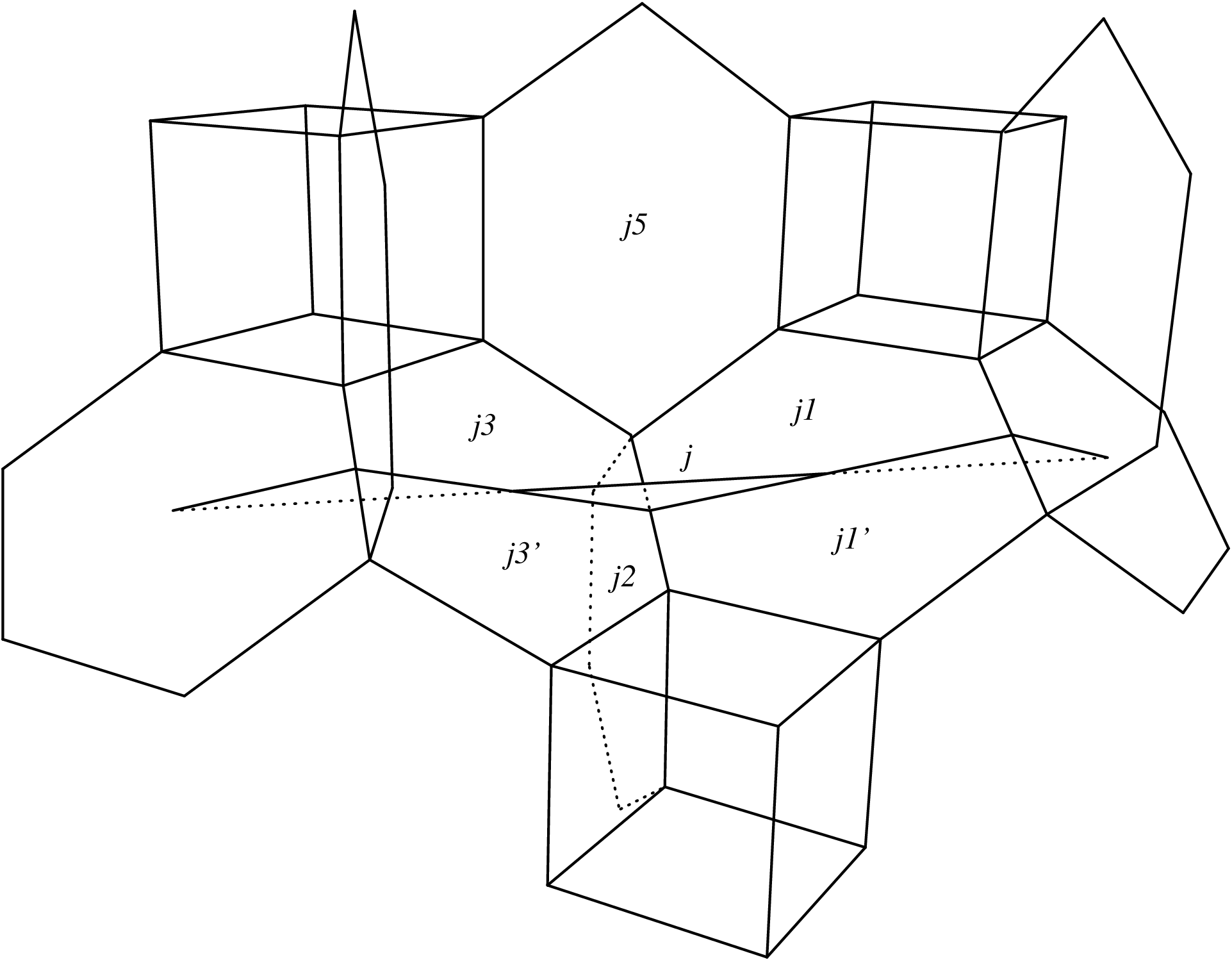}

Using this complex, we can describe the spin foam sum of the two Polyakov loops as follows. It is given by 
\be
\label{spinfoamsumWilsonloopTprimedual}
\b \tr_j U_{C_1} \tr_j U_{C_2}\ket = \frac{1}{Z}\,\sum_{F\;|\;\pa F = (\Ct_1\cup \Ct_2,j)}
\left(\prod_{f\subset\kt} (2j_f+1)\right) 
\left(\prod_{v\subset\kt} A_v\right)
\left(\prod_{f\subset \kappa}\;(-1)^{2j_f}\,\e^{-\frac{2}{\beta}\,j_f(j_f + 1)}\right)\,.
\ee
The difference to \eq{spinfoamsumpartitionfunction} consists of the modification of the complex and the boundary condition $\pa F = (\Ct_1\cup \Ct_2,j)$.
The boundary condition $\pa F = (\Ct_1\cup \Ct_2,j)$ requires that the spin on the loop edges is $j$. The attachement of triangles along $\Ct_1\cup \Ct_2$ creates 
two types of new vertices in the complex: vertices in the middle of hexagons along $\Ct_1\cup \Ct_2$, and vertices in the middle of the boundary edge between such hexagons. In the first case, the vertex amplitude is trivial, i.e.\
\be
A_v = 1\,.
\ee
To the second type of vertex we associate a tetrahedral spin network whose edges and vertices correspond to faces and edges around this vertex: \setlength{\jot}{0.5cm}
\bea
A_v &=& (-1)^{j_3 - j'_3}\,(-1)^{j_1 - j'_1}\,(-1)^{j_1 + j_3 + j_2 + j}\;\;\parbox{3.8cm}{\includegraphics[height=2cm]{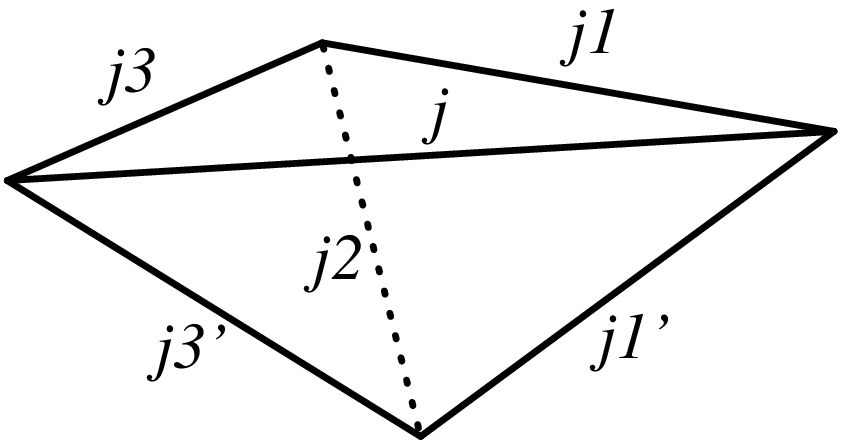}} \\
&=& (-1)^{j_3 - j'_3}\,(-1)^{j_1 - j'_1}\,(-1)^{j_1 + j_3 + j_2 + j}\,\sixj{j_1}{j_3}{j_2}{j'_3}{j'_1}{j}\,.
\label{newvertexamplitude}
\eea
The spins $j_1$, $j_2$ and $j_3$ are read off from one of the two vertices not adjacent to $j$: if the edge with spin $j$ is drawn at the top (as in \fig{Tdualprime}), this vertex is on the left side of $j$ in the direction of passage of the Polaykov loop, i.e.\ on the left side in the direction from $j_3$, $j'_3$ towards $j_1$, $j'_1$. 
\setlength{\jot}{0cm}

\section{Worldsheet interpretation of spin foams}
\label{worldsheetinterpretationofspinfoams}

\subsection{Definition of worldsheets}

To arrive at the worldsheet interpretation of spin foams, we have to apply a further modification to the complex $\kt$.
We ``frame'' $\kt$, so that it becomes a 3-complex. Under this framing each 2--cell $f$ of $\kt$ is turned into a 3-cell $f'$ that has the topology
of $f\times (0,1)$. Neighbouring cells are connected as in \fig{framing} and \fig{projection}a. The resulting 3--complex is called $\kp$. 
The precise metric properties of $\kp$ do not matter as long as it has the required cell structure. The framing of $\kt$ induces also a framing of the boundary $\kt$. Each 1--cell $e\subset\pa\kt$ is thickened into a 2--cell $e'$ that has the topology of a half--open strip $[0,1]\times (0,1)$. Note that the boundary $\pa e'$ of $e'$ is disconnected. When we speak of the boundary $\pa\kp$ of $\kp$, we mean the union of all such framed edges $e'$: they form two ribbons---the framed version of the two loops $\Ct_1$ and $\Ct_2$ (see \fig{Tdualprime}).

\pic{framing}{Under the framing three faces of $\kt$ along an edge become three 3--cells that intersect along 2--cells.}{6cm}{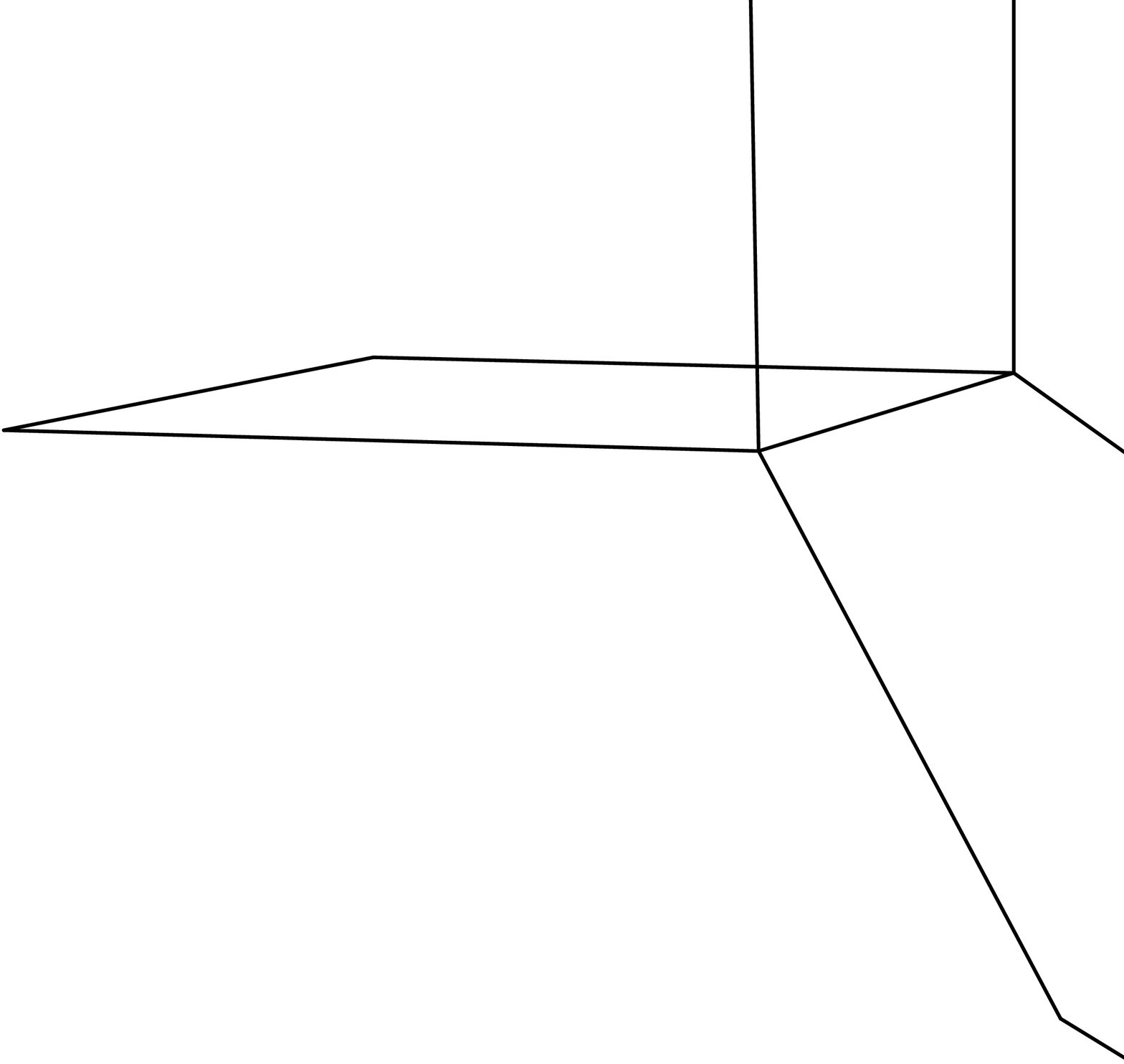}

Consider a compact embedded surface\footnote{The embedding implies, in particular, that surface does not intersect with itself.} $S$ in $\kp$ whose boundary lies in $\pa\kp$. Take a framed 3--cell $f'$ in $\kp$ and determine the intersection $S\cap \pa f'$ of the surface with the cell boundary $\pa f'$. In general, this intersection can be empty or consist of loops, lines and points. The cell boundary $\pa f'$ has the topology of an open annulus, so there are two types of loops: loops that are contractible in $\pa f'$ and loops that are not. 
Let us assume that for any $f'\subset\kp$, the intersection $S\cap \pa f'$ contains only loops of the non--contractible kind.
We count the number of such loops in $\pa f'$ and call it $N_f$. Obviously, this number does not change if we apply a homeomorphism to $S$ that is connected to the identity and maps cell boundaries $\pa f'$ onto themselves. In this limited sense, the numbers $N_f$, $f\subset\kt$, are topological invariants. 

Moreover, they satisfy constraints. To see this, consider a triple $f_1, f_2, f_3$ of faces that intersect along an edge $e$ of $\kt$. Correspondingly, we have three framed faces $f'_1, f'_2, f'_3$ of $\kp$ that intersect along 2--cells $e'_{12}, e'_{23}, e'_{31}$ (see \fig{framing}). The surface $S\subset\kp$ induces non--contractible loops within the boundaries $\pa f'_1, \pa f'_2, \pa f'_3$ (see \fig{inducedloops}). Clearly, each loop in a boundary $\pa f'_i$ borders exactly one loop from another boundary $\pa f'_j$, $i\neq j$. This pairing of loops implies that the numbers $N_{f_1}, N_{f_2}, N_{f_3}$ satisfy the triangle inequality
\be
|N_{f_1} - N_{f_2}| \le N_{f_3} \le N_{f_1} + N_{f_2}\,.
\ee
If we write $j_f = N_f /2$, this is precisely the spin coupling constraint that defines a spin foam $F$ with spins $j_f$.
We see therefore that the numbers $N_f$ define spin foams $F$ on $\kt$! We will show, in fact, that for every spin foam $F$ there is a surface $S$ whose loop numbers are given by $F$, and if we restrict the surfaces suitably there is a bijection between surfaces in $\kp$ and spin foams on $\kt$.
On the boundary this relation induces a correspondence between curves on $\pa\kp$ and spin networks on $\pa\kt$.

We will first define a suitable class of surfaces and curves, and then prove that the bijection holds. Motivated by the well--known conjectures about gauge--string dualities, we call the surfaces and curves worldsheets and strings. Equivalence relations will be furnished by homeomorphisms $h:\Lambda\to\Lambda$ on $n$--complexes $\Lambda$, $n = 2,3$, that 
\begin{enumerate}
\item map boundaries $\pa c$ of $n$--cells $c$ onto themselves, and
\item are connected to the identity through homeomorphisms with property 1. 
\end{enumerate}
Let $\mathrm{Homeo}(\Lambda)$ denote the set of such restricted homeomorphisms.  
\begin{definition}
A string $\gamma$ on $\kp$ is an embedded, not necessarily connected, compact closed curve in the boundary of $\kp$ where
for each 2--cell $c$ of $\pa\kp$ the intersection $\gamma\cap c$ consists of lines and each line intersects $\pa c$ in two end points that are not contractible in $\pa c$.
\end{definition}
We consider two strings $\gamma$ and $\gamma'$ as equivalent if they are related by a homeomorphism $h\in\mathrm{Homeo}(\pa\kp)$.
\begin{definition}
A worldsheet $w$ on $\kp$ is an embedded, not necessarily connected, compact surface in $\kp$ such that
\begin{itemize}
\item[(i)] the boundary $\pa w$ is a string on $\pa\kp$, and
\item[(ii)] for each 3--cell $f'$ of $\kp$ the intersection $w\cap f'$ consists of disks and each disk intersects $\pa f'$ in a loop that is non--contractible in $\pa f'$.
\end{itemize}
\end{definition}
We consider two worldsheets $w$ and $w'$ as equivalent if they are related by a homeomorphism $h\in\mathrm{Homeo}(\kp)$. 

\pic{inducedloops}{A surface $S$ induces loops in the boundary of 3--cells of $\kp$.}{8cm}{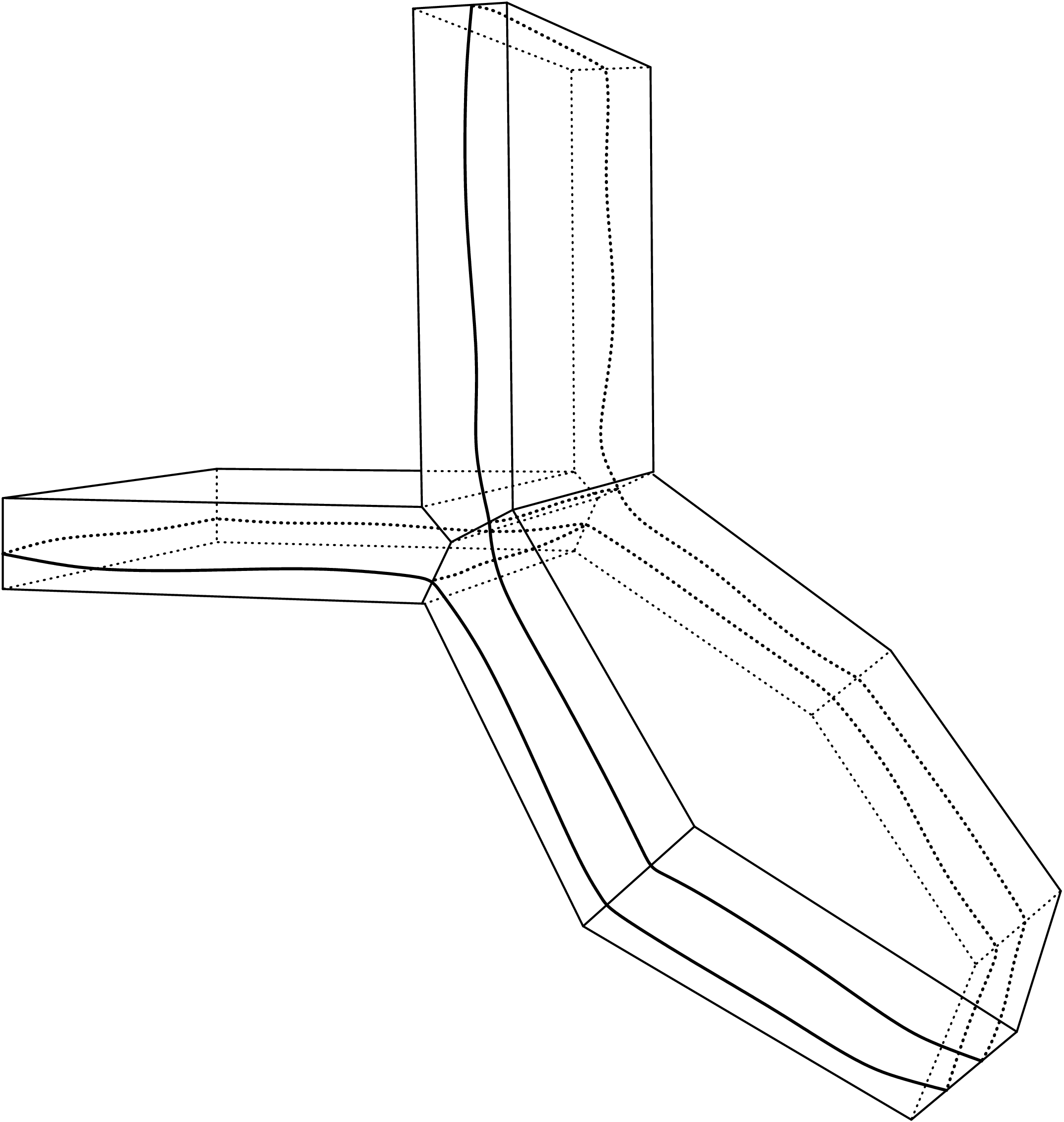}

\subsection{Correspondence between spin foams and worldsheets}
\label{correspondencebetweenspinfoamsandworldsheets}

Since the boundary of $\kp$ has the topology of $S^1\cup S^1$, the correspondence between strings on $\pa\kp$ and spin networks on $\pa\kt$ 
is rather trivial. It is clear from the definition that a string on $\pa\kp$ is a union of $N_1$ disjoint loops along $\Ct_1\times (0,1)$ and $N_2$ disjoint loops along $\Ct_1\times (0,1)$. We denote this string by $\gamma_{\Ct_1,N_1}\cup\gamma_{\Ct_2,N_2}$. On the other hand, the only possible spin networks are given by the loops $(\Ct_1,j_1)\cup (\Ct_2,j_2)$ with spin $j_1$ and $j_2$. Therefore, a one--to--one correspondence is set up by asssociating the string $\gamma_{\Ct_1,2j_1}\cup\gamma_{\Ct_2,2j_2}$ to the spin network $(\Ct_1,j_1)\cup (\Ct_2,j_2)$.  

Let us now choose fixed values for the spins $j_1$ and $j_2$. Denote the set of all spin foams $F$ s.t.\ $\pa F = (\Ct_1,j_1)\cup (\Ct_2,j_2)$ by $\clF$, and let $W$ stand for the set of worldsheets s.t.\ $\pa w = \gamma_{\Ct_1,2j_1}\cup\gamma_{\Ct_2,2j_2}$.
\begin{proposition}
There is a bijection $f: \clF \to W$ between spin foams $F$ on $\kt$ s.t.\ $\pa F = (\Ct_1,j_1)\cup (\Ct_2,j_2)$ and worldsheets $w$ on $\kp$ s.t.\ $\pa w = \gamma_{\Ct_1,2j_1}\cup\gamma_{\Ct_2,2j_2}$.
\end{proposition}
\begin{proof}
We start by constructing a map $f: \clF \to W$. Then, we will show that $f$ is injective and surjective.

Let $F$ be a spin foam in $\clF$. 
Consider the vertices $v$ of $\kp$ where six 3--cells intersect. Denote the set of these vertices as $V'$.
Around each vertex $v\in V'$ we choose a closed ball $B_{\epsilon}(v)$ of radius $\epsilon$. The intersection of the balls with cells of $\kp$ defines a new, finer complex that we call $\kappa'_{\pm}$. We can view this complex as the union of two complexes $\kappa'_+$ and $\kappa'_-$, where $\kappa'_+$ results from $\kappa'_\pm$ by removing the interior of all balls $B_{\epsilon}(v)$:
\be
\kappa'_+ = \kappa'_{\pm} \backslash \bigcup_{v\in V'}  B^\circ_{\epsilon}(v)
\ee 
$\kappa'_-$, on the other hand, is the subcomplex of $\kappa'_\pm$ that remains when we keep the balls $B_{\epsilon}(v)$ and delete the rest.
Every 3--cell $f'$ of $\kp$ is a union
\be
f' = f'_+ \cup \bigcup_i f'_{-i}
\ee
where $f'_+$ is a 3--cell of $\kappa'_+$ and the $f'_{-i}$, $i=1,\ldots,n$, are 3--cells in $\kappa'_-$.

In order to construct the worldsheet corresponding to the spin foam $F$, we will first build a surface in the complex $\kappa'_+$.
In the second step, we will also fill the balls $B_{\epsilon}(v)$ with surfaces, so that the union of all surfaces gives a worldsheet in $\kp$.
Consider an arbitrary face $f$ of $\kt$ with spin $j_f$ determined by the spin foam $F$. The corresponding 3--cell $f'_+$ in $\kappa'_+$ has the topology of a closed 3--ball with two punctures. Its boundary  $\pa f'_+$ is an open annulus. In each such 3--cell $f'_+$ we place $N_f = 2j_f$ disjoint closed disks whose boundary is given by non--contractible loops in $\pa f'_+$. Along every edge $e$ in the interior of $\kappa'_+$ three 3--cells $f'_{+1}, f'_{+2}, f'_{+3}$ intersect. Due to the spin coupling conditions, the numbers $N_{f'_{+1}}, N_{f'_{+2}}, N_{f'_{+3}}$ satisfy the triangle inequality
\be
|N_{f'_{+1}} - N_{f'_{+2}}| \le N_{f'_{+3}} \le N_{f'_{+1}} + N_{f'_{+2}}\,.
\ee
This implies that we can rearrange the disks in such a way that their boundary edges are pairwise congruent at the shared boundaries of the cells $f'_{+1}, f'_{+2}, f'_{+3}$. We repeat this procedure for every edge $e\subset \kappa'_+$ where three 3--cells meet, and thereby obtain a compact embedded surface $w_+$ in $\kappa'_+$. Up to homeomorphisms $h\in\mathrm{Homeo}(\kappa'_+)$, this surface is uniquely determined by our procedure.

We now explain how we fill the ``holes'' in $\kappa'_+$, so that we get a surface in the entire complex $\kp$. 
Each ball $B_{\epsilon}(v)$ defines a subcomplex of $\kappa'_-$ as depicted in \fig{projection}a. It consists of six 3--cells $c_1,\ldots, c_6$. The boundary $\pa c_i$ of each cell is topologically an open annulus, and subdivided into five 2--cells. Four of these 2--cells are shared with neighouring 3--cells $c_j$, $j\neq i$, and one of them lies in the boundary $\pa B_{\epsilon}(v)$ of the ball. We call the former type of 2--cell internal, and the latter one external.

To fill this complex with surfaces, it is helpful to use another, topologically equivalent complex that is shown in \fig{projection}b: the interior of the ball $B_{\epsilon}(v)$ corresponds to the interior of a tetrahedron and the boundary $\pa B_{\epsilon}(v)$ is projected onto one of the four triangles. This triangle has three punctures. 

\psfrag{1}{$\sst 1$}
\psfrag{2}{$\sst 2$}
\psfrag{3}{$\sst 3$}
\psfrag{4}{$\sst 4$}
\psfrag{5}{$\sst 5$}
\psfrag{6}{$\sst 6$}
\begin{figure}
\setlength{\unitlength}{1cm}
\begin{center}
(a)\quad\parbox{4cm}{\includegraphics[height=4cm]{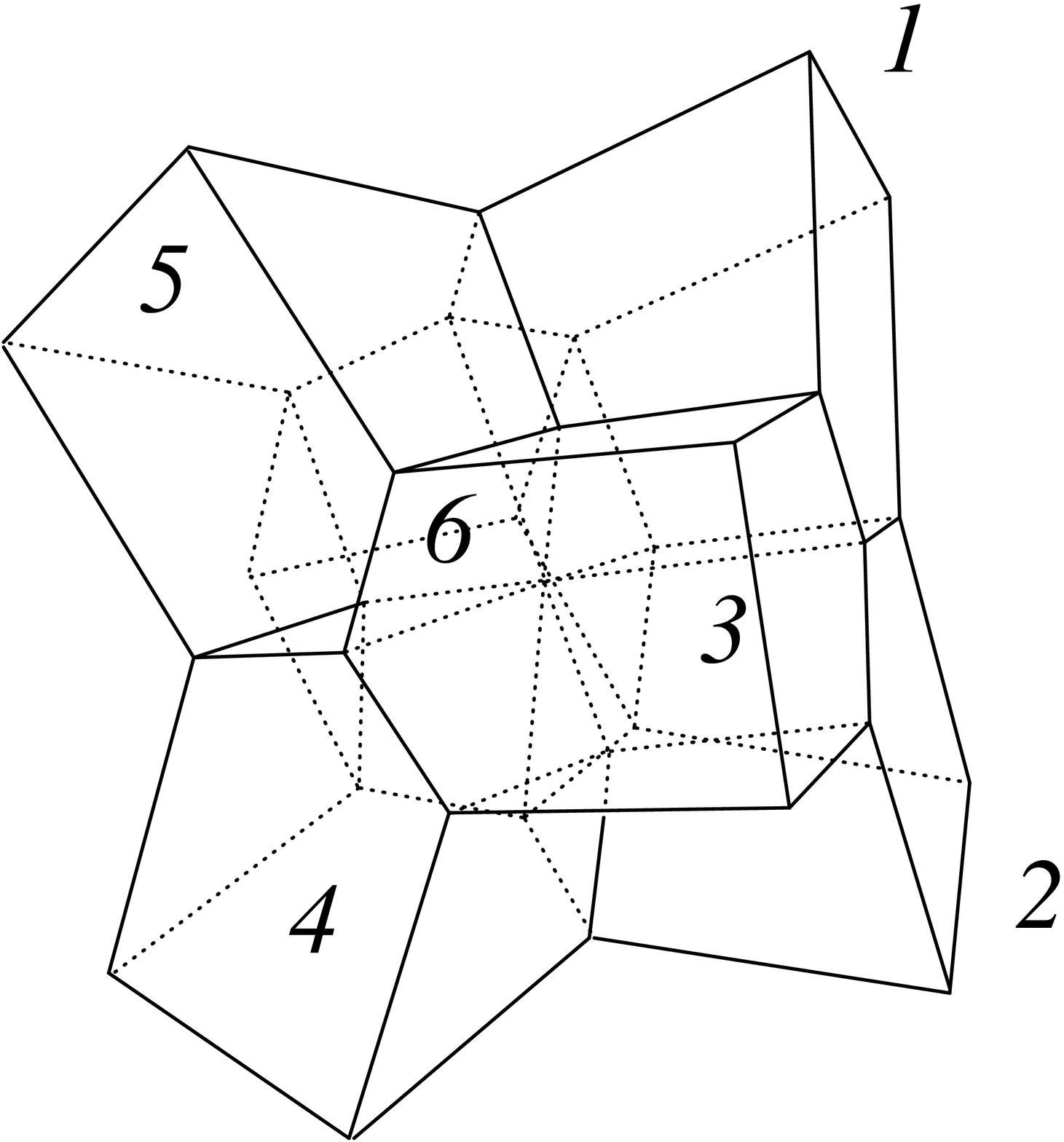}}
\hspace{1cm}
(b)\quad\parbox{4cm}{\includegraphics[height=5.5cm]{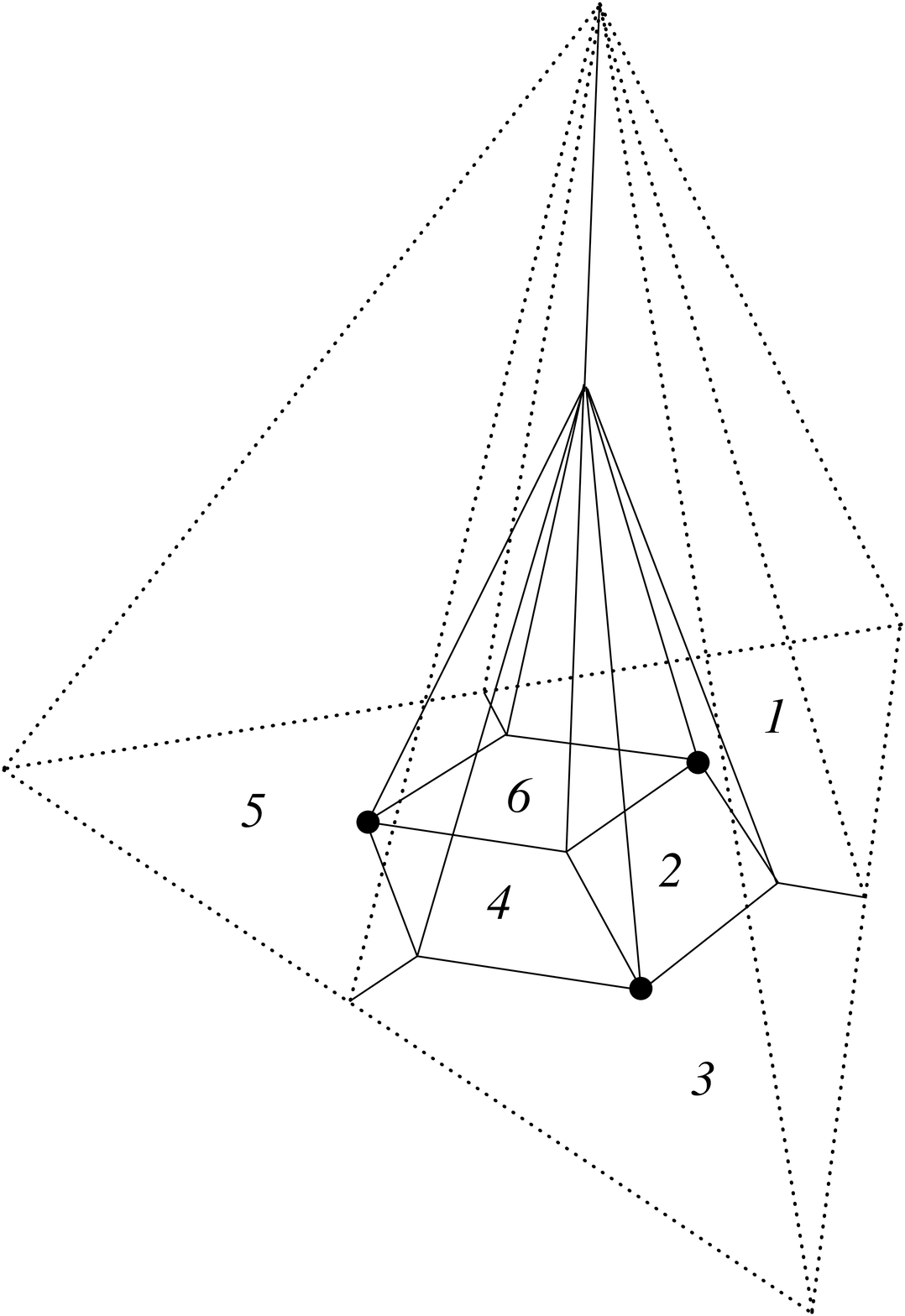}}
\end{center}
\caption{\label{projection} 
(a) A closed ball $B_\epsilon(v)$ in $\kp$ around a vertex $v$ where six framed cells meet. The cells of $\kp$ induce a cell structure in the ball. The resulting cell complex is topologically equivalent to the complex in \fig{projection}b. (b) A tetrahedron in $\bR^3$ with an open triangle at the bottom, and triangles removed on the three other sides. The boundary of the ball $B_\epsilon(v)$ is mapped onto the bottom triangle. Solid lines delineate the boundaries between 3--cells in the interior of the tetrahedron. The three thick dots indicate punctures. The three missing triangles in the boundary form a fourth puncture.}
\end{figure}

For every ball $B_{\epsilon}(v)$, the surface $w_+$ induces an embedded closed curve $\gamma_v$ along the boundary $\pa B_{\epsilon}(v)$. The curve consists of $n$ loops $l_i$, i.e.\ $\gamma_v = l_1\cup \cdots\cup l_n$. In the alternative representation of \fig{projection}b, the curve appears as a set of embedded loops in the bottom triangle that wind around the three punctures (see \fig{loopsatbottom}). To create the surface in $B_{\epsilon}(v)$, we will cover the $n$ loops by $n$ disks in $B_{\epsilon}(v)$. This will be done in such a way that condition (ii) for worldsheets is satisfied.

\pic{loopsatbottom}{Example of an induced loop in the boundary of the ball $B_{\epsilon}(v)$.}{7cm}{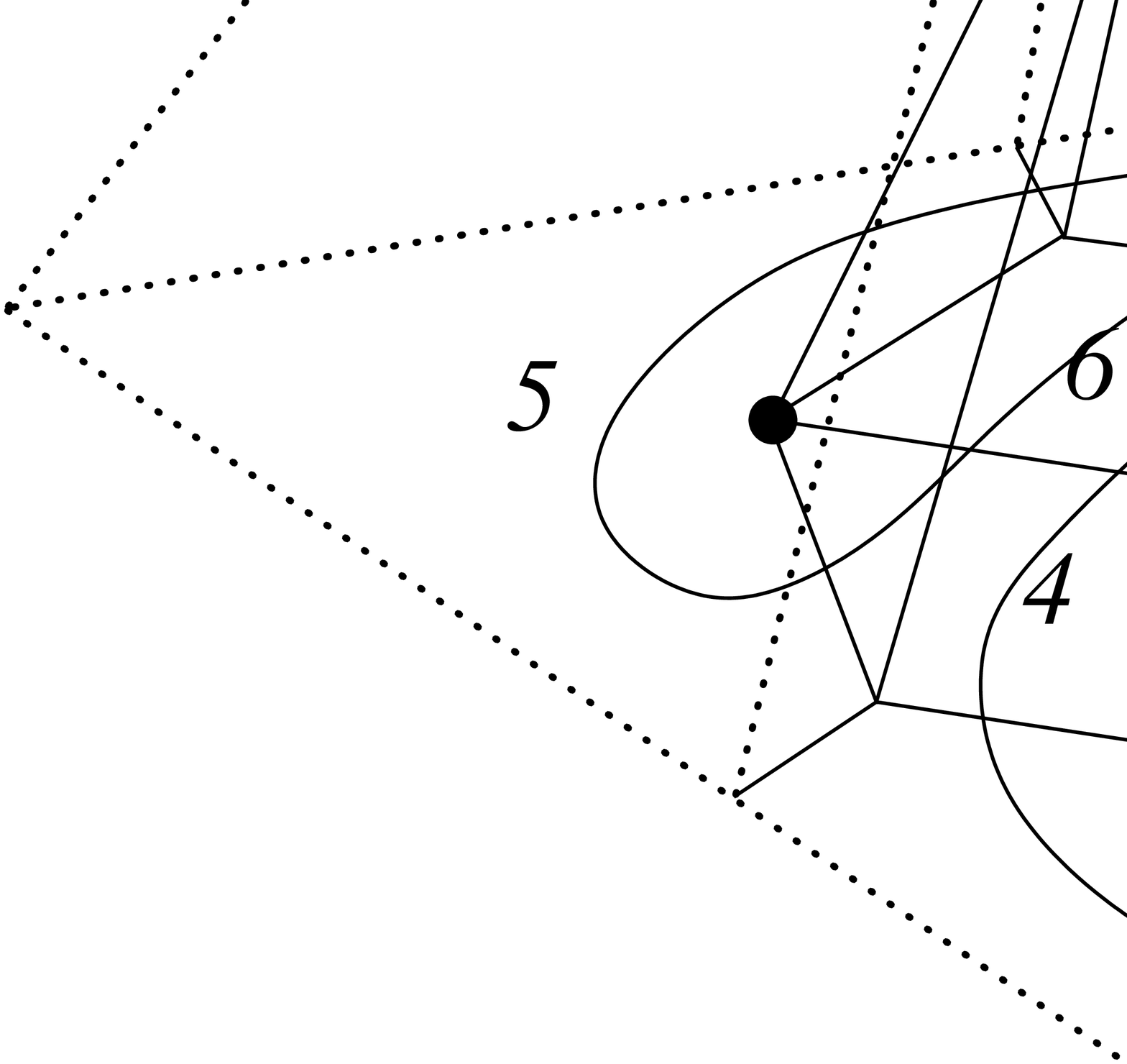}

Consider a single 3--cell $c_i$ in the ball $B_{\epsilon}(v)$, and the one external 2--cell in its boundary $\pa c_i$. The intersection of the curve $\gamma_v$ with this 2--cell gives a number of lines $e_{ik}$, $k=1,\ldots K_i$. Each of the two end points of a line $e_{ik}$ is located on a 1--cell shared by an external and an internal 2--cell of $\pa c_i$. Let us now draw a line from one of the end points through the internal 2--cell to the vertex in the center of $B_{\epsilon}(v)$. The same is done for the second end point. Together with the original line $e_{ik}$, these lines form a loop in the cell boundary $\pa c_i$. We fill this loop with a disk $d_{ik}$ in $c_i$, so that the intersection $d_{ik}\cap c_i$ is again the loop. 

We repeat this procedure for every line $e_{ik}$ in the cell $c_i$, and then in every cell $c_i$. Along the boundary between neighbouring 3--cells, we glue the disks together: when a line $e_{ik}$ is connected to another line $e_{i'k'}$, $i\neq i'$, the corresponding disks are glued together along the internal 2--cell $\pa c_i \cap \pa c_{i'}$. This can be done in such way that the resulting total surface intersects only in one point: at the vertex in the center of $B_{\epsilon}(v)$, like a stack of sheets that are pinched together. Let us call this surface $w_-$. 

Observe that $w_-$ satisfies property (ii) in the subcomplex $B_{\epsilon}(v)$. Due to the way we have placed disks outside of $B_{\epsilon}(v)$, every line $e_{ik}$ connects 1--cells of $\pa B_{\epsilon}(v)$ that are disconnected. As a result, each loop $d_{ik}\cap c_i$ is non--contractible in $\pa c_i$. 

To arrive at an embedded surface, we need to remove the point of degeneracy at the center of the ball $B_{\epsilon}(v)$. We do so by moving the different parts of $w_-$ slightly apart, and in such a way that no new components are created in the intersections $w_- \cap c_i$. The latter ensures that the new surface $w_-$ still has property (ii). Up to homeomorphisms $h\in\mathrm{Homeo}(B_{\epsilon}(v))$ which leave $\gamma_v$ invariant, $w_-$ is the unique embedded surface that is bounded by $\gamma_v$ and meets condition (ii). 

After filling each ball $B_{\epsilon}(v)$ with such a surface $w_-$, we take the union of the surfaces $w_-$ with $w_+$. This gives us an embedded compact surface $w$ in $\kp$. Let us check if $w$ meets requirement (i) and (ii) of the definition of a worldsheet. 

Due to the arrangement of disks in 3--cells $c$ of $\kappa'_+$, the induced loops in the boundary $\pa c$ never connect 1--cells that are connected in $\pa c$. This means, in particular, that the induced curve in the boundary $\pa\kp$ consists of lines in 2--cells $c$, where each line connects two disconnected 1--cells of $\pa c$. Therefore, the boundary of each line cannot be contracted in $\pa c$, and the surface $w$ has property (i). 

How about property (ii)? The surface has the desired property for the cells of $\kappa'_+$, and we showed the same for each subcomplex $B_{\epsilon}(v)$. 
It is clear from this that $w$ has property (ii) in $\kp$. 

We conclude that $w$ is a worldsheet on $\kp$. The whole construction defines a map $f: \clF \to W$ from spin foams to worldsheets.

Next we prove that $f$ is injective and surjective. Let $F$ and $F'$ be two different spin foams. There must be a face $f\subset\kt$ for which $N_f \neq N'_f$. 
This implies that the corresponding worldsheets $w$ and $w'$ are different, since they have different invariants under the homeomorphisms $h\in \mathrm{Homeo}(\kp)$. Thus, $f$ is injective. 

To check surjectivity, consider an arbitrary worldsheet $w\in W$. Within each 3--cell $c$ of $\kp$, the worldsheet induces $N_f$ disks that are bounded by non--contractible loops in $\pa c$. The numbers $N_f$ define a spin foam $F$ with spins $j_f = N_f/2$. From $F$ we construct another worldsheet $w' = f(F)$. Provided the balls $B_{\epsilon}(v)$ are chosen small enough, the intersections $w\cap\kappa^+$ and $w'\cap\kappa^+$ are related by a homeomorphism $h\in\mathrm{Homeo}(\kp_+)$. Inside the balls $B_{\epsilon}(v)$, the worldsheet $w'$ consists of disks that are bounded by loops in $\pa B_{\epsilon}(v)$. Up to homeomorphisms $h_v\in\mathrm{Homeo}(B_{\epsilon}(v))$ that leave $\gamma_v$ invariant, there is precisely one way to cover these loops by disks in $B_{\epsilon}(v)$ such that property (ii) is met. For sufficiently small $\epsilon$, the intersection $w\cap B_{\epsilon}(v)$ has property (ii) as well, and must be related to $w'\cap B_{\epsilon}(v)$ by a homeomorphism $h_v\in\mathrm{Homeo}(B_{\epsilon}(v))$. 
Thus, there is a homeomorphism $h\in\mathrm{Homeo}(\kp)$ that relates $w$ and $w'$, and $w' = f(F) = w$. This shows that $f$ is surjective and completes the proof.
\end{proof}

\section{String representation of 3d SU(2) lattice Yang--Mills theory}
\label{stringrepresentationof3dSU2latticeYangMillstheory}

By using the correspondence between spin foams and worldsheets, we can now translate the exact dual representations \eq{spinfoamsumpartitionfunction} and \eq{spinfoamsumWilsonloopTprimedual} into exact string representations of 3d SU(2) Yang--Mills theory.
The string representation is defined on a complex $\kp$ that arises from a framing of the 2--skeleton of a tesselation $\kt$ by cubes and truncated rhombic dodecahedra (see \fig{TandTdual}, \fig{framing} and \fig{projection}a). Under the framing, faces $f$ of the 2--skeleton become 3--cells $f'$ of the framed complex $\kp$. Vertices $v$ turn into vertices $v'\subset\kp$, where six framed 3--cells $f'$ intersect. The set of these vertices $v'$ is denoted by $V'$. The 3--cells $f'$ of $\kp$ belong to two groups: 3--cells $f'$ that originate from square faces $f$ of $\kt$ (and correspond to faces $f\in\kappa$), and those arising from hexagonal faces in $\kt$. Worldsheets and strings are defined as certain surfaces and curves in the framed complex $\kp$ (see \sec{worldsheetinterpretationofspinfoams}).

With these conventions, the partition function is given by a sum over closed worldsheets:
\be
\label{stringrepresentationpartitionfunction}
Z = 
\sum_{w\;|\;\pa w = \emptyset}
\left(\prod_{f'\subset\kp} (N_{f'} + 1)\right) 
\left(\prod_{v'\subset V'} A_{v'}(\{N_{f'}/2\})\right)
\left(\prod_{f\subset\kappa}\;(-1)^{N_{f'}}\,\e^{-\frac{1}{2\beta}\,N_{f'}(N_{f'} + 2)}\right)
\ee
The amplitude has three contributions: every framed face contributes with a factor $N_{f'} + 1$, where $N_{f'}$ is the number of components of the worldsheet in $f'$. In addition, square faces give an exponential and a sign factor $(-1)^{N_{f'}}$. For each vertex $v\in V'$, we get a $6j$--symbol
\be
A_{v'}(\{N_{f'}/2\}) = \sixj{N_{f'_1}/2}{N_{f'_2}/2}{N_{f'_3}/2}{N_{f'_4}/2}{N_{f'_5}/2}{N_{f'_6}/2} 
\ee
where the $f'_i$ are the six 3--cells that intersect at $v'$.

For the expectation value of two Polyakov loops $(C_1,j)$ and $(C_2,j)$ (as defined in section \sec{subsectionPolyakovloops}), an additional modification of the 2--skeleton was required: we insert a sequence of triangles along two loops $\Ct_1$ and $\Ct_2$ (see \fig{Tdualprime}). Under the framing, the two loops become ribbons. The expectation value of the Polyakov loops is equal to a sum over worldsheets that are bounded by $2j$ strings along the first ribbon $\Ct_1\times (0,1)$ and by $2j$ strings along the second ribbon $\Ct_2\times (0,1)$. Denoting these strings as $\gamma_{\Ct_1,2j}\cup \gamma_{\Ct_2,2j}$, the sum takes the form
\be
\label{stringrepresentationPolyakovloops}
\b \tr_j U_{C_1} \tr_j U_{C_2}\ket =
\sum_{w\;|\;\pa w = \gamma_{\Ct_1,2j}\cup \gamma_{\Ct_2,2j}}
\left(\prod_{f'\subset\kp} (N_{f'} + 1)\right) 
\left(\prod_{v'\subset V'} A_{v'}(\{N_{f'}/2\})\right)
\left(\prod_{f\subset\kappa}\;(-1)^{N_{f'}}\,\e^{-\frac{1}{2\beta}\,N_{f'}(N_{f'} + 2)}\right)\,.
\ee
The difference to \eq{spinfoamsumpartitionfunction} consists of the modification of the complex and the boundary condition $\pa w = \gamma_{\Ct_1,2j}\cup \gamma_{\Ct_2,2j}$. The attachement of triangles along $\Ct_1\cup \Ct_2$ creates two types of new vertices in $\kp$: vertices in the middle of framed hexagons along the ribbons, and vertices in the middle of the boundary between such hexagons. In the first case, the vertex amplitude is trivial, i.e.\
\be
A_{v'} = 1\,.
\ee
To the second type of vertex we attribute a factor\renewcommand{\arraystretch}{1.5} 
\be
A_{v'} = (-1)^{(N_{f'_3} - N'_{f'_3})/2}\,(-1)^{(N_{f'_1} - N'_{f'_1})/2}\,(-1)^{(N_{f'_1} + N_{f'_3} + N_{f'_2} + 2j)/2}\,
\left\{
\begin{array}{ccc}
N_{f'_1}/2 & N_{f'_3}/2 & N_{f'_2}/2 \\ 
N'_{f'_3}/2 & N'_{f'_1}/2 & j
\end{array}
\right\}
\ee
The labelling is analogous to the labelling by spins in eq.\ \eq{newvertexamplitude}. 

In this string representation, $N$--ality dependence and string breaking take on a very concrete form.
For spin $j=1/2$, the boundary string consists of two loops $\gamma_{\Ct_1,1}$ and $\gamma_{\Ct_2,1}$: one along the ribbon $\Ct_1\times (0,1)$ and the other one along the ribbon $\Ct_2\times (0,1)$. Since every worldsheet has to be bounded by the string $\gamma_{\Ct_1,1}\cup\gamma_{\Ct_2,1}$, there is necessarily a connected component of the worldsheet that connects the boundary strings $\gamma_{\Ct_1,1}$ and $\gamma_{\Ct_2,1}$. The string between quarks is ``unbroken''. When we go to $j=1$, on the other hand, we have a pair $\gamma_{\Ct_1,2}$ of strings along $\Ct_1\times (0,1)$ and a pair $\gamma_{\Ct_1,2}$ of strings along $\Ct_2\times (0,1)$. In this case, the four single strings can be either connected by two surfaces that go across the distance between the Polaykov loops, or each pair is connected to itself by a tube--like surface. In the latter case, the string between quarks is ``broken''. As we go to higher spins, the worldsheet can consist of several extended surfaces, several tube--like surfaces or a mixture of both.

\section{Discussion}
\label{discussion}

In this paper, we showed that 3d SU(2) lattice Yang--Mills theory can be cast in the form of an exact string representation.
Our starting point was the exact dual (or spin foam) representation of the lattice gauge theory. We demonstrated that spin foams can be equivalently described as self--avoiding worldsheets of strings on a framed lattice. This lattice arose in two steps: we replaced the original cubic lattice by a tesselation, where at every edge only three faces intersect. Then, we took the 2--skeleton of this complex, and framed (or thickened) it by choosing an open neighbourhood of it in $\bR^3$. We proved that there is a bijection between a subset of surfaces in the framed complex and spin foams in the unframed complex. This allowed us to translate the partition function from a sum over spin foams into a sum over closed worldsheets. The expectation value of two Polyakov loops with spin $j$ became a sum over worldsheets that are bounded by $2j$ strings along each loop.

To our knowledge, this is the first example of an exact and fully explicit string representation of SU(2) lattice Yang--Mills theory in three dimensions\footnote{In the case of 2d QCD, an exact string representation was found by Gross and Taylor \cite{GrossTaylor2dQCD1,GrossTaylor2dQCD2}.}. Not surprisingly, it differs from a simple Nambu--Goto string. When a worldsheet does not run more than once through faces (i.e.\ when $N_{f'}\le 1$), the $6j$--symbols in the amplitude become trivial and the exponent in \eq{stringrepresentationpartitionfunction} is proportional to the area of the worldsheet. In these cases, the weighting resembles that of the Nambu--Goto string. In general, however, a worldsheet intersects several times with the same cell, and then we have an interaction due to nonlinear dependences on $N_{f'}$. That is, in addition to interactions by merging and splitting, there is an interaction of directly neighouring strings. Note that this does not preclude the possibility that a Nambu--Goto string gives a good effective description in special cases or regimes.
 
It is interesting to compare this result to the AdS--CFT correspondence, where the gauge--string duality is constructed by completely different methods.
One should also observe the difference between our ``non--abelian'' worldsheets and the surfaces that arise in abelian lattice gauge theory. In the case of U(1), the theory can be transformed to a sum over closed 2--chains, and in this sense one has a sum over surfaces. The worldsheets of our string representation are of the same type as long as $N_{f'}\le 1$. When the occupation number increases, however, the surfaces can be ``jammed'' against each other along faces without being ``added'' like abelian 2--chains.

At a practical level, the present worldsheet picture could be useful for analyzing the dual representation. It could be helpful, for example, when thinking about ``large'' moves in Monte Carlo simulations \cite{ChristensenCherringtonKhavkine}: by inserting an entire worldsheet into a given worldsheet, one can create a non--local change in spin foams that is compatible with the spin--coupling conditions.

A possible shortcoming of the present work is the restriction on the shape of surfaces. It was needed in order to establish the bijection between worldsheets and spin foams. From a mathematical perspective, it would be more elegant to admit arbitrary compact self--avoiding surfaces, and to characterize spin foams as certain topological invariants. We hope to obtain such a characterization in future work.

\begin{acknowledgments}
We thank Wade Cherrington, Dan Christensen, Alejandro Perez and Carlo Rovelli for discussions. This work was supported in part by the NSF grant PHY-0456913 and the Eberly research funds.
\end{acknowledgments} 

\bibliography{bibliography}
\bibliographystyle{hunsrt}  

\end{document}